\documentclass[aps,twocolumn,prl,superscriptaddress,showpacs,preprintnumbers,amsmath,amssymb]{revtex4}

\usepackage{graphicx}
\usepackage{color}

\begin{document}

%usepackage{graphicx}%
%\usepackage{dcolumn}
%\usepackage{amsmath}
%\usepackage{color}

%\newcommand{\hl}[1]{\textcolor{red}{#1}}

%\begin{document}

\title{Fabrication of high-temperature quasi-two-dimensional superconductors at the interface of a ferroelectric Ba$_{0.8}$Sr$_{0.2}$TiO$_{3}$ film and an insulating parent compound of La$_{2}$CuO$_{4}$}
\author{Dmitrii~P.~Pavlov}
\affiliation{Zavoisky Physical-Technical Institute, FRC KazanSC of
RAS, 420029 Kazan, Russia}

\author{Rustem~R.~Zagidullin}
\affiliation{Zavoisky Physical-Technical Institute, FRC KazanSC of
RAS, 420029 Kazan, Russia}

\author{Vladimir~M.~Mukhortov}
\affiliation{Southern Scientific Center of RAS,
%
%Chehova 41,
%
344006
Rostov-on-Don, Russia}

\author{Viktor~V.~Kabanov}
\affiliation{Zavoisky Physical-Technical Institute, FRC KazanSC of
RAS, 420029 Kazan, Russia} \affiliation{Department for Complex
Matter, Jozef Stefan Institute, 1000 Ljubljana, Slovenia}

\author{Tadashi~Adachi}
\affiliation{Department of Engineering and Applied Sciences, Sophia
 University, Tokyo, Japan}

\author{Takayuki~Kawamata}

\author{Yoji Koike}
\affiliation{Department of Applied Physics, Tohoku University,
Sendai, Japan}

\author{Rinat~F.~Mamin}
\affiliation{Zavoisky Physical-Technical Institute, FRC KazanSC of
RAS, 420029 Kazan, Russia}

\date{\today}

%\maketitle

%\begin{abstract}
%

\begin{abstract}
%{\bf \noindent
{ We report the first observation of superconductivity in
heterostructure consisting of an insulating ferroelectric film
(Ba$_{0.8}$Sr$_{0.2}$TiO$_3$) grown on an insulating parent compound
of La$_2$CuO$_4$ with [001] orientation. The heterostructure was prepared
by magnetron sputtering on a non-atomically-flat surface with
inhomogeneities of the order of 1-2 nm. The measured superconducting
transition temperature T$_c$ is about 30K.
We have shown that superconductivity is confined near the
interface region. Application of a weak magnetic field perpendicular
to the interface leads to the appearance of the finite resistance. That
confirms the quasi-two-dimensional nature of the superconductive
state. The proposed concept promises ferroelectrically controlled
interface superconductivity which offers the possibility of novel
design of electronic devices.}
%}

\end{abstract}

%%% PACS numbers

\maketitle

Up to now the creation of  high-T$_c$ quasi-two-dimensional
superconductivity (HTq2DSC)~\cite{c01,c02,c03,c04,c01_2,c02_2} as
well as a quasi-two-dimensional electron gas
(q2DEG)~\cite{c05,c06,c07,c08,c09,c10,c11} at the interface were
impossible without tailoring the atomically perfect
interfaces~\cite{c01,c02,c03,c04,c05,c06,c07,c08,c09,c10,c11,c12}.
The realization of HTq2DSC area is a long-term goal because of
potential applications~\cite{c13,c14} and the possibility to study
quantum phenomena in two dimensions~\cite{c07,c15,c16}. Typical
approaches to the realization of quasi-two-dimensional
superconducting layer rely on creation of an ``ultrathin'' film of a
known superconductor~\cite{c13,c14}. However it is important not
only to get HTq2DSC, but also to have the ability to control
superconducting states by magnetic and electric fields. In this
Letter we present the experimental realization of HTq2DSC by
increasing the carrier concentration in a thin layer of the parent
compound of the high temperature superconductor (PCHTSC)  at the
interface with the ferroelectric. By simple consideration the
additional current carriers at the interface occur due to
electrostatic potential arising from polar discontinuity. It allows
to change the conduction properties of the heterostructures by
switching the polarization in the ferroelectric. This approach
allows to get heterostructures with relatively simple technology
because the requirements for boundary condition are less stringent.

\par
Tailoring q2DEG and HTq2DSC at the interface is impossible without
the deep understanding of the nature of quasi-two-dimensional
states. First, the q2DEG has been created at the heterointerfaces
between two insulating oxides, LaAlO$_3$/SrTiO$_3$~\cite{c03}, and
unique transport properties were observed owing to strong electronic
correlations~\cite{c05,c06,c07,c08,c09,c10,c11}. In this case the
system becomes superconducting below 300 mK~\cite{c07}. Then the
superconductivity at 30 K in bilayers of an insulator
(La$_2$CuO$_4$) and a metal (La$_{1.55}$Sr$_{0.45}$CuO$_4$), neither
of which is superconducting in isolation, was
reported~\cite{c01,c02}. The interplay of superconductivity with the
antiferromagnetic order was also studied in the metal-insulator
cuprates superlattices~\cite{c01_2,c02_2}. The price for these
result in both cases was that the interfaces should be atomically
perfect~\cite{c01,c07,c12}. In the second case it was
considered~\citep{c01,c11}, that the interface must be atomically
perfect to obtain the superconductivity at the interface in copper
oxides, because the coherence length is very short ($\xi$=1-3
nm)~\cite{c17}. In the case of a ferroelectric oxide deposited on
the copper oxide, the conditions are not so stringent for the
appearance of the effect: inhomogeneities of the order of $\xi$ are
possible if their envelope is much greater than $\xi$. Thus in this
Letter we report the first observation of superconductivity in
heterostructure consisting of an insulating ferroelectric film
(Ba$_{0.8}$Sr$_{0.2}$TiO$_3$) on an insulator single crystal
(La$_2$CuO$_4$). Here the results had been obtained on the
heterostructure, created by relatively simple method of magnetron
sputtering and using more simple conditions for the interface. We
show experimentally that it is possible to get q2DEG on  a
non-atomically-flat interface. And we obtain superconductivity with
T$_c$=30 K at the heterointerfaces between two insulating oxides,
which is a hundred times higher than T$_c$ in LaAlO$_3$/SrTiO$_3$
heterostructure. We would like to underline that using a
ferroelectric oxide in the heterostructures allows us to fabricate
the interface of two insulating oxides with different structure of
elementary cells and to have more simple RF-sputtering method for
tailoring the heterostructure. Moreover, using a ferroelectric
material as an upper layer of the heterostructure brings interesting
new physics, which opens the possibility to change the properties of
the heterostructures by switching the polarization in the
ferroelectric layer.

\begin{figure}[tb]
%\centering
%\includegraphics[angle=-90,width=1.0\linewidth]{Fig_1ab}
\includegraphics[angle=0,width=1.0\linewidth]{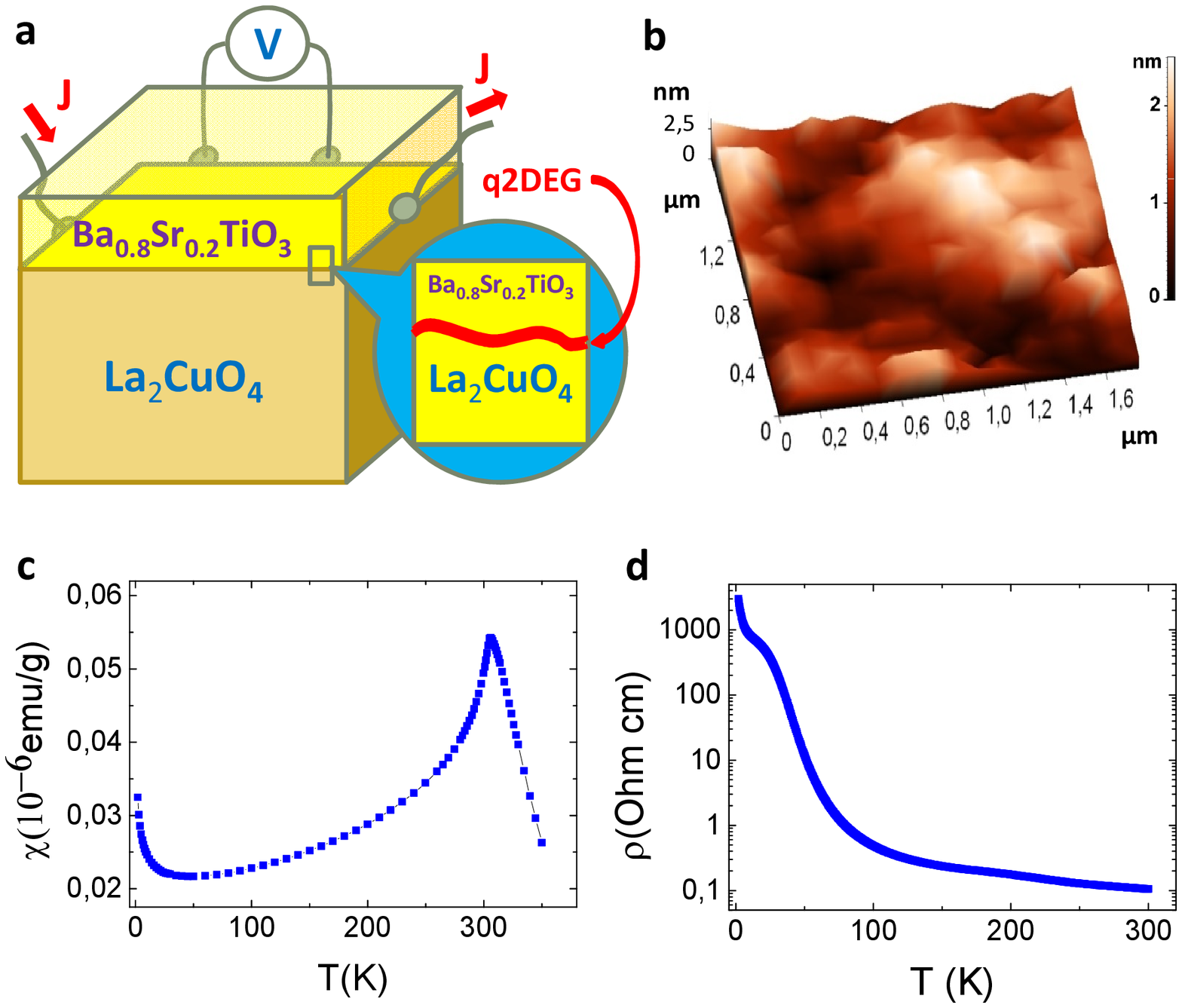}
% \vspace{-5.0cm}
%
\caption{ The schematic structures of
Ba$_{0.8}$Sr$_{0.2}$TiO$_3$/La$_2$CuO$_4$ (a) with q2DEG (shown in
%transparent
red); AFM image of the La$_2$CuO$_4$ single crystal
surface without the film (b) illustrates the inhomogeneity of the
interface. The temperature dependence of the magnetic susceptibility
(c), and the temperature dependence of the resistivity (d) of
La$_2$CuO$_4$ single crystal (without ferroelectric film). }
\label{fig:AC}
\end{figure}

\begin{figure}[htb]
\includegraphics[angle=0,width=0.85\linewidth]{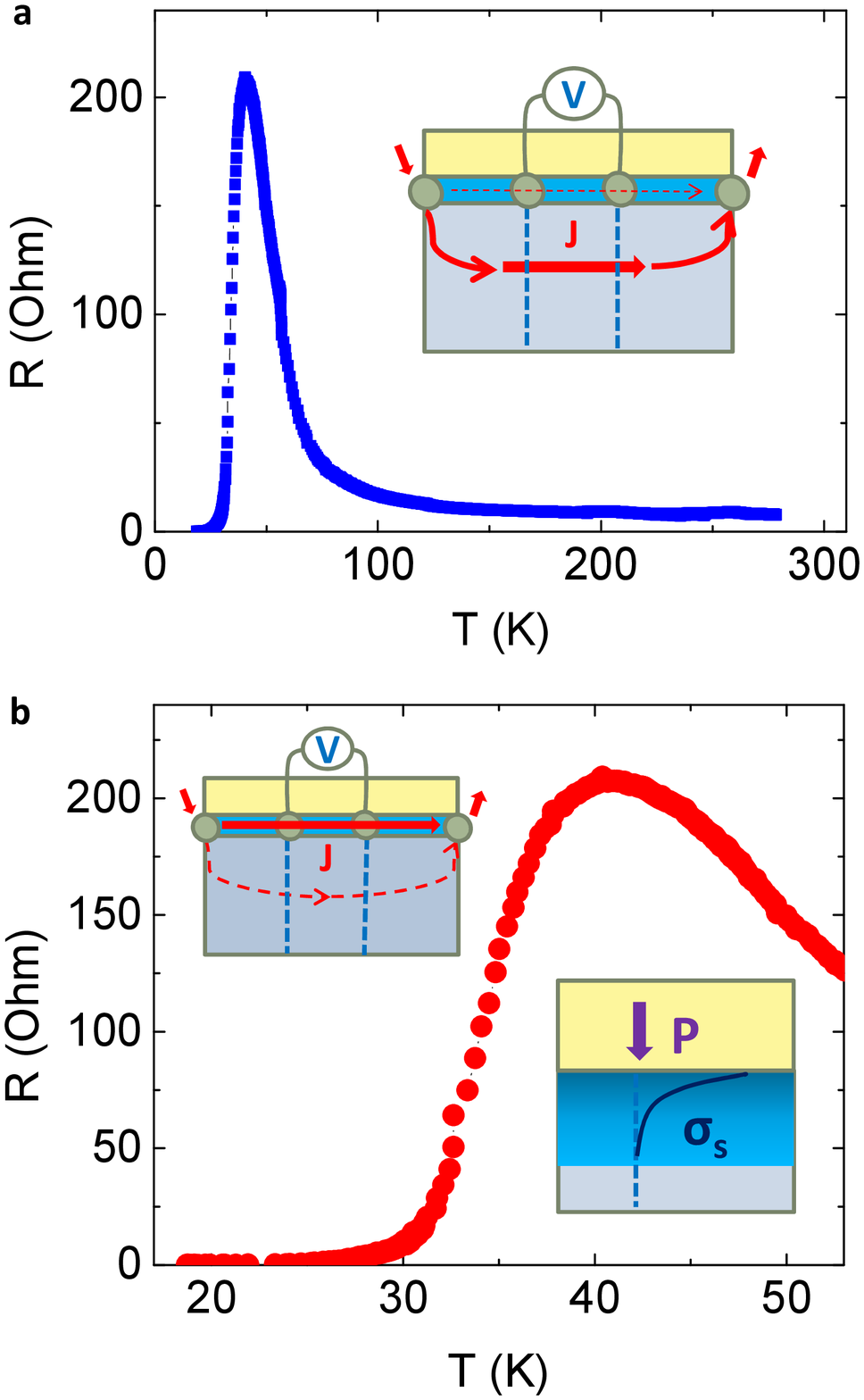}
\caption{ The temperature dependence of the resistance of
Ba$_{0.8}$Sr$_{0.2}$TiO$_3$/La$_2$CuO$_4$ heterostructure in the
wide temperature range (a) and at low temperatures (b) (the
results of the same measurements).   }
\label{fig:ZF}
\end{figure}

\begin{figure}[htb]
%\centering
%\vspace{-2.5cm}
\includegraphics[angle=0,width=1.00\linewidth]{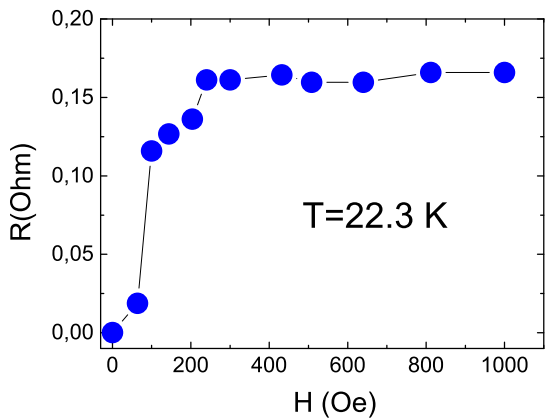}
\caption{ The magnetic field dependence of the resistance of
Ba$_{0.8}$Sr$_{0.2}$TiO$_3$/La$_2$CuO$_4$ heterostructure.   }
 \label{fig:Phase-diagramm}
\end{figure}

In our investigation, a La$_2$CuO$_4$ (LCO) single crystal was grown
using a traveling-solvent-floating-zone technique and was
characterized by magnetic susceptibility and resistivity
measurements. And then a Ba$_{0.8}$Sr$_{0.2}$TiO$_3$ (BSTO)
ferroelectric oxide was deposited on $ab$ surface of the single
crystal (see Fig. 1a) by reactive sputtering of stoichiometric
targets using RF plasma (RF-sputtering) method \cite{c18} at 650 C
and partial pressure of oxygen of 0.7 Torr (details of the deposition
are in the Supplemental Material \cite{supplementary}).
Therefore, we tried to combine the advantages of both approaches
described above in order to get the superconducting properties of the
interface  in the easy way.
We have used a LCO single crystal as a substrate in order to obtain
a high T$_c$. We use a relatively simple method of creating the
interface, and typical surface roughness of LCO single crystal
determined from atomic force microscopy data before deposition is
about 1-2 nm (with the size in the plane of approximately 200-300
nm, see Fig. 1b), which is slightly more than one unit cell in c
direction (1.3 nm in LCO). Heteroepitaxial BSTO ferroelectric thin
film (thickness of 200 nm from atomic force microscopy data) was
deposited on LCO single crystal (001) substrate. BSTO belongs to
ferroelectric perovskites. In the ferroelectric phase below the
Curie temperature of the ferroelectric phase transition (T$_c$ = 353
K)~\cite{c19} it has a tetragonal unit cell. The as-grown film shows
built-in polarization in the [001] crystallographic direction which
was determined by the x-ray measurements. The
temperature dependences of magnetic susceptibility $\chi$(T) and
resistivity $\rho$(T) of the LCO single crystal are shown in Fig. 1c
and 1d. The peak in $\chi$(T) clearly observed around 306 K
corresponds to the N\'eel temperature below which a long-range
antiferromagnetic order is formed. The temperature dependence of
resistivity $\rho$(T) (Fig. 1d) is usual for LCO~\cite{c20,c21}.
Both of these results are typical for this system~\cite{c20,c21} and
indicate a good quality of the crystal.

Resistance measurements on the interface of the heteroctructure ware
performed by four contacts method. The electrodes were applied by
using silver paste on the LCO surface at the boundary with film as
schematically  shown on Fig. 1a. The electrodes were in contact with
the interface. The distance between potential electrodes was
different in the different experiments. The distribution of the
current, flowing by different routes at different temperatures,
depends on relation between the substrate conductance and the interface
conductance. At high temperature the main current flows through the
substrate. Below 50 K the current flows mainly in the interface region. The
temperature dependence of the resistance of
Ba$_{0.8}$Sr$_{0.2}$TiO$_3$/La$_2$CuO$_4$ heterostructure in the
wide temperature range (see Fig. 2a) shows that above ~40 K the
resistance has usual semiconducting behavior. At low temperatures
(Fig. 2a and Fig. 2b) the resistance drops very rapidly and
superconducting behavior is observed. Thus the interface between the
ferroelectric and insulating oxides shows superconducting behavior
with a high T$_c$ of about 30 K (Fig. 2). The beginning of the transition
to the superconducting state occurs around 40 K, similar to what is
observed in bulk La$_{2-x}$Sr$_x$CuO$_4$ (LSCO) single crystals at
optimal doping~\cite{c20,c21}. When a weak magnetic field is applied
to the heterostructure in the direction perpendicular to the surface
of the interface, the finite resistance of the interface appears and
it increases with the increasing of the field (see Fig. 3) as it was
predicted~\cite{c22}. The magnetic field was applied perpendicular
to the surface and parallel to c axis of the LCO substrate at T=22.3
K. The magnetic field dependence of the heterostructure resistance
shows that non-zero resistance appears at a very low field. The
H$_{c1}$ for a thin layer of superconductor is very small and the
magnetic field penetrates in the superconducting layer.
In that case the system demonstrates flux-flow resistance. That
confirms a quasi-two-dimensional nature of the superconductive
state (see also Supplemental Material \cite{supplementary}.) We did not perform
the measurements in higher magnetic fields
intentionally, since we know from the previous experience~\cite{c23}
that the effects of magnetostriction in relatively small magnetic
fields can lead to partial peeling of the film from the substrate
resulting in partial or complete disappearance of the observed
effect. On the other hand, we believe that in our case, by analogy
with LSCO, H$_{c2}$ is of the order of 29-81 T, which is
inaccessible for us.

The most common mechanism for q2DEG is the polarization catastrophe
(PC) model~\cite{c15,c10}, which was also discussed for the case of
ferroelectric/dielectric interface~\cite{c22,c23,c24,c25,c26}. The
polar discontinuity at the interface leads to the divergence of the
electrostatic potential.  In order to minimize the total energy, it
is necessary to shield the electric field arising from this. As a
consequence, both the lattice system and the energy spectrum of the
current carriers are restructured~\cite{c22}, and the increase of
the current carriers density occurs in a narrow interface area. This
occurs in a self-consistent manner, so that rearranging the energy
spectrum of the carriers and increasing their concentration in the
interface region leads to the formation of a narrow metal region
near the interface on the part of the LCO as shown in the right
insert of Fig. 2b. Our estimates show that if we assume that the
polarization of the ferroelectric is P = 30 $\mu$K/cm$^2$ (it gives
$\sigma_S$=1,875\ 10$^{14}$ 1/cm$^2$) and the screening length in
LCO is $d_{Sc}$ = 0.45 nm, then the concentration corresponding to
the doping level, at which the superconducting state is observed in
La$_{2-x}$Sr$_{x}$CuO$_4$ ($x$ = 0.05-0.26), will be achieved in a
narrow region of LCO in the second-third interface layers of the
CuO$_2$ planes (details are presented in the Supplemental Material \cite{supplementary}).

In addition to this possibility, the occurrence of HTq2DSC is
possible due to the impact of cation interdiffusion (primarily Ba or
Sr from BSTO to LCO) and oxygen non-stoichiometry. Barium or
strontium diffusion is unlikely due to low diffusion coefficient at
650 C~\cite{c27}. Reduction of oxygen during deposition of the film
is also unlikely, since the process is carried out at elevated
oxygen pressure. For that matter, an introduction of additional
oxygen in this process would be more likely. But the following three
experimental facts argue against this. The first is that q2DEG was
created at the interface of the
Ba$_{0.8}$Sr$_{0.2}$TiO$_3$/LaMnO$_3$ heterostructure~\cite{c23}. It
would be unlikely that a change in the oxygen concentration in
LaMnO$_3$ could lead to the appearance of q2DEG, because it was
shown experimentally for this case that the occurrence of q2DEG is
related to the direction of polarization in the ferroelectric, and
arises only in the case of polarization directed perpendicular to
the interface~\cite{c23}.
\begin{figure}[htb]
\includegraphics[angle=0,width=0.95\linewidth]{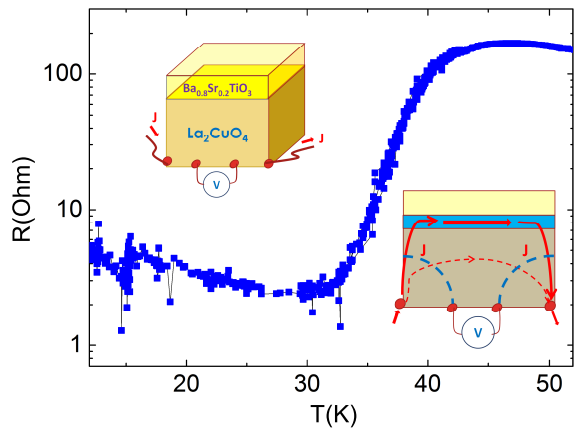}
\caption{  The temperature dependence of the resistance of
Ba$_{0.8}$Sr$_{0.2}$TiO$_3$/La$_2$CuO$_4$ heterostructure from the
substrate side.  The temperature dependence of the heterostructure
resistance is measured by electrodes deposited on the
La$_{2}$CuO$_4$ surface opposite to the surface with the film. }
\label{fig:superfluid_common}
\end{figure}
Our sample was obtained by the same technology using the same
equipment. The second is that the application of the magnetic field,
which partially destroys the contact at the interface between
Ba$_{0.8}$Sr$_{0.2}$TiO$_3$ and LaMnO$_3$, leads to q2DEG
disappearance. From this fact it was concluded that the occurrence
of q2DEG is related to the proximity effect, rather than to
diffusion processes. And the third fact is illustrated in Fig. 4.
Here we applied electrodes for resistance measurements on the back
side (single crystal side) of the heterostructure, and electrodes
were not in contact with the interface (see
upper-left insert in Fig. 4). In this case the superconducting state is
not observed directly. The resistance decreases below a certain
temperature but not below 4 Ohm.
We believe that the current line distributions are strongly
different at different temperatures (see lower-right insert in Fig.
4) and depend on the relation between conductance of the substrate
and the interface. At high temperature the main current flows
through substrate. Below 50 K the current flows mainly in the
interface region, and the resistance shows qualitatively the same
behavior as the resistance measured with the electric contacts
at the top (Fig. 2b, see also Supplemental Material \cite{supplementary}). But superconductivity is
not observed directly because the surface of substrate is not
superconducting. It means that the oxygen does not penetrate the
surface layer during the film deposition. The possibility of the
reduction of oxygen in the interface area during deposition had been
also discussed for the case of bilayers
La$_2$CuO$_4$/La$_{1.55}$Sr$_{0.45}$CuO$_4$~\cite{c01,c02}. It was
concluded that ``Interstitial oxygen in La$_2$CuO$_{4+\delta}$ is
mobile and, in particular in very thin films, it diffuses out of the
sample on the scale of hours or days''~\cite{c01}.

All this findings strongly indicate that we observed superconductivity
in heterostructures consisting of an insulating ferroelectric film
(Ba$_{0.8}$Sr$_{0.2}$TiO$_3$) grown on an insulating single crystal with
[001] orientation (La$_2$CuO$_4$).
This heterostructure was created by magnetron sputtering
on a non-atomically-flat surface. Our results open a new page in
creating interfaces with q2DEG and HTq2DSC, since it has been shown
experimentally that it is possible to create HTq2DSC by a relatively
simple method at the boundary of the ferroelectric and parent
compound of the high temperature superconductor (PCHTSC). We believe
that these results will have large impact in the field and will be
interesting for a broad scientific community, because a large number of
new heterostructures may be fabricated by this technique, and a large
number of different groups may use this method. Note, that with this
technique we obtain superconductivity with T$_c$=30 K, which is a hundred
times higher than T$_c$ in LaAlO$_3$/SrTiO$_3$ heterostructures.

In conclusion q2DEG is formed at the interface, which becomes
HTq2DSC state when the temperature is lowered below 30 K. The
HTq2DSC arises from strongly increasing carrier density localized in
the interface area in copper oxide while the polar discontinuity at
the interface leads to the divergence of the electrostatic potential
due to the polarization catastrophe~\cite{c05,c10}. This allows
to control the interface superconductivity by applying an electric
field, as it was done in the case of the ionic liquid~\cite{c28}. It
opens the possibilities to use these phenomena in a novel design of
electronic devices.

We thank N.N. Garif'yanov, R. Khasanov (PSI), A.V.Leontyev, Yu.I.
Talanov and R.V. Yusupov for help and advices. We acknowledge I.A.
Garifullin and V.E. Kataev for discussions. R.F.M. is grateful to
his wife Aigyul for stimulation of this research. V.V.K.
acknowledges financial support from Slovenian Research Agency
Program P1-0040.The work at ZPhTI FRC KazanSC RAS was funded by
Russian Scientific Foundation, research project No. 18-12-00260.

{\bf Supplemental Material.}
\ \\

{\bf I. Sample preparation and characterization.}
\ \\

The La$_2$CuO$_4$ (LCO) single crystal was grown using a
traveling-solvent-floating-zone technique and was characterized by
magnetic susceptibility and resistivity measurements. The surface of
the single crystal was polished using diamond paste up to roughness
of 1-2 nm and a size of the roughness in the plane was 200-300 nm
(Fig. 1 and Fig. 1c, 1d of the article). Heteroepitaxial
Ba$_{0.8}$Sr$_{0.2}$TiO$_3$ (BSTO) thin film was deposited on LCO
(001) substrate by RF-sputtering.
%
% of stoichiometric polycrystalline
%target.
%
Transparent BSTO film was realized by means of
layer-by-layer growth (Frank-van der Merwe mechanism).
The growth of single-crystal films occurs from the dispersed oxide
phase, which is formed in the plasma of a high-frequency discharge
when a ceramic target of the same compound is sputtered at a cluster
level. The applied RF power made it possible to spray the composite
oxide at the level of a cluster that served as the vapor phase for
the growing film. The substrate temperature was 650 C, it was kept constant 
during the deposition of the films and the oxygen pressure
was 0.7 Torr. This method of film growth has been repeatedly tested
and is constantly used by us to obtain films of this composition
[1-3]. The layered structure of the growth was confirmed by studying
of the films by electron and atomic force microscopy. The average
layer roughness of the films was ~ 0.4 nm. All details of the growth
conditions have been previously reported in Ref.[1-4]. The
orientation of the film was determined by the x-ray measurements. It
was found that c-axis of the film is perpendicular to the interface.
The sample was examined by atomic force microscopy method after the
deposition of the film. The inhomogeneity of a film surface was
10-20 nm with the size in the plane of approximately 200-300 nm
(Fig. 1b of the article). The measured thickness was 200 nm (Fig.
2a, 2b).  The as-grown film consists of 180-degrees domains which
are not compensated due to an interfacial strain and show built-in
polarization in the [001] crystallographic direction[1,2]. The size
of domains was about 200 nm (Fig. 2a).

The results of the deposition of BSTO films on different substrate 
suggests that the diffusion does not take place [1-3]. The most diffusive 
component in BSTO compound is oxygen. Therefore we repeat the 
experiments on a La$_2$CuO$_4$ single crystal (from the same batch) heated in the 
deposition chamber, at the same oxygen partial pressure and temperature, 
used during the heterostructure fabrication. After that we measured the 
resistivity of this La$_2$CuO$_4$ single crystal. We did not observe any changes in 
the temperature dependence of resistivity. Superconducting transition was not 
observed and the resistivity showed the same semiconducting behaviour up to low 
temperatures as in the samples before heating in the oxygen atmosphere (see also the results of the susceptibility measurements in section IV).

\begin{figure}[htb]
%\centering
% \vspace{-1.0cm}
 \includegraphics[angle=0,width=1.0\linewidth]{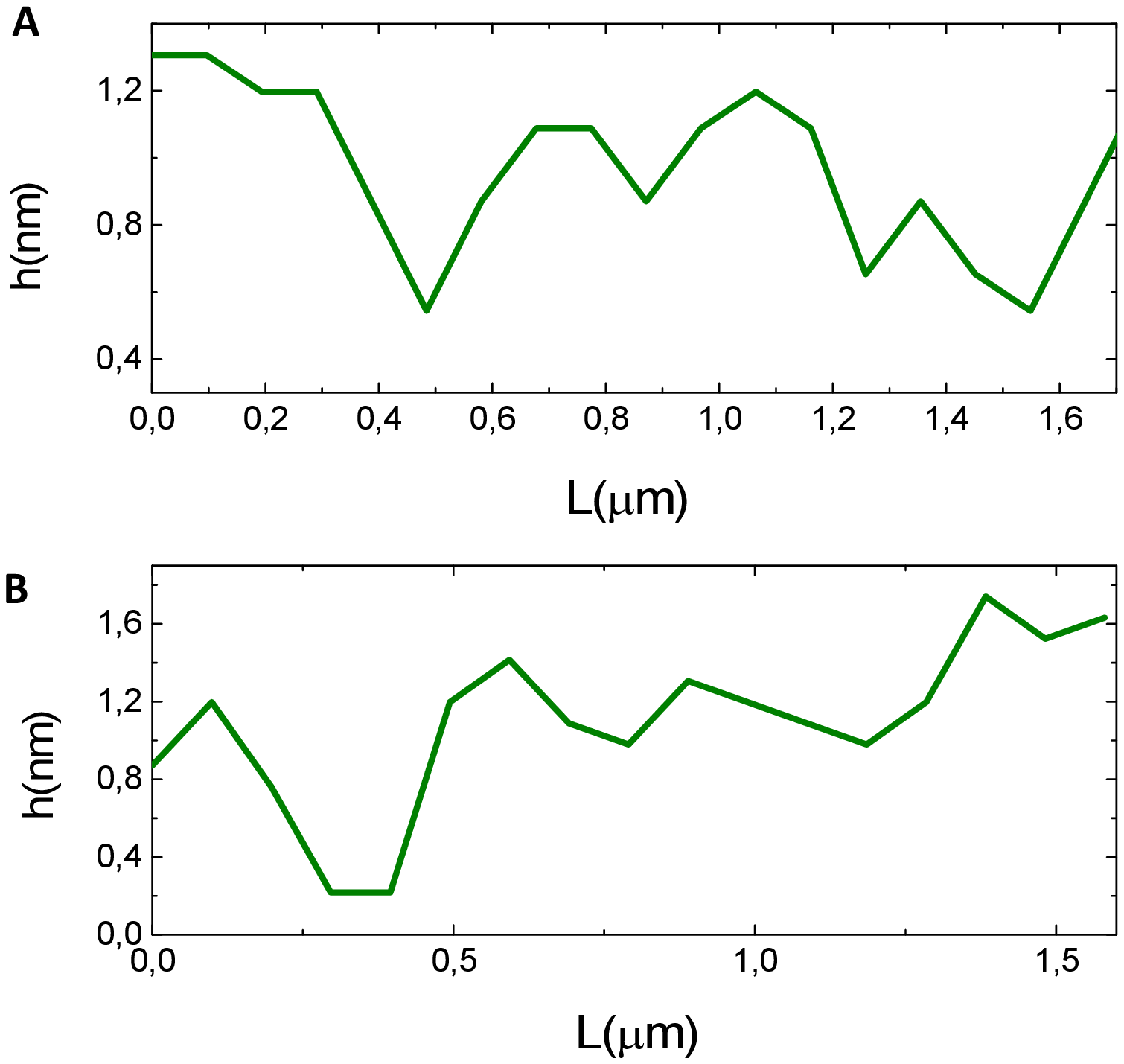}

 \vspace{-0.0cm}
% closed and the open symbols refer to FeSe$_{0.94}$ and
\caption{ The surface of the La2CuO4
single crystal. The cross-section of surface profile of the single
crystal without the film at "good" polish (A) and "bad" polish (B)
areas are shown. }
 \label{fig:Phase-diagramm}
\end{figure}

\begin{figure}[htb]
%\centering
% \vspace{-1.0cm}
 \includegraphics[angle=0,width=1.0\linewidth]{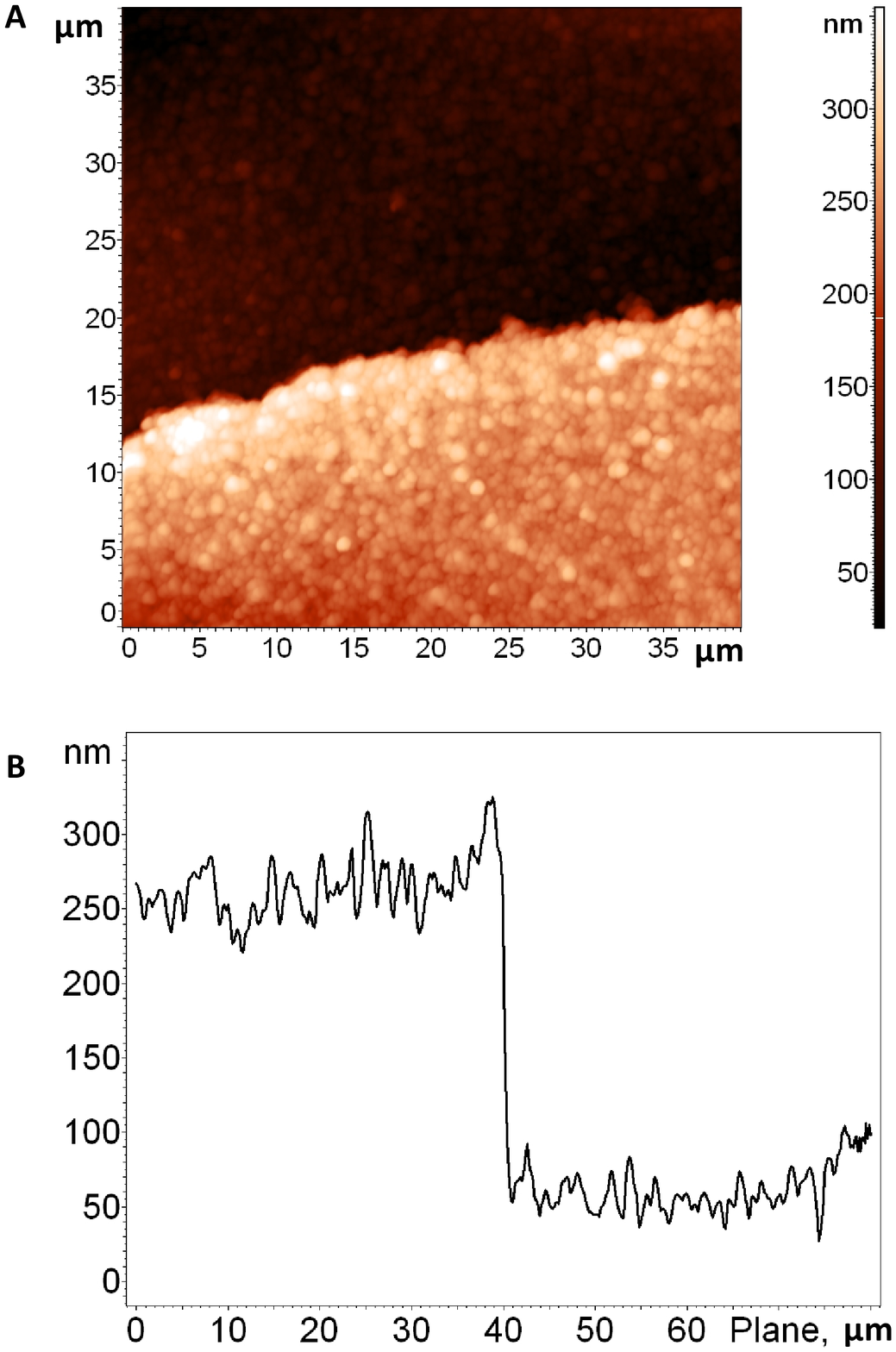}
%\includegraphics[width=1.0\linewidth]{Fig_3}
%\includegraphics[angle=-90,width=1.0\linewidth]{Fig_3}
%\includegraphics[angle=-90,width=1.0\linewidth]{111}
% \vspace{-0.0cm}
\caption{ The film thickness measurement.
AFM profile (A) and cross-section (B) of surface of the
Ba$_{0.8}$Sr$_{0.2}$TiO$_3$/La$_2$CuO$_4$ heterostructure in the
area of film boundary show the thickness of about 200 nm.  }
 \label{fig:Phase-diagramm}
\end{figure}

%\begin{figure}[htb]
% \vspace{-0.00cm}
%\includegraphics[angle=0,width=1.0\linewidth]{Fig_Sp3.eps}
% \vspace{-0.0cm}
%\caption{Figure 3 of Supplementary. The temperature dependence of
%the resistance of Ba$_{0.8}$Sr$_{0.2}$TiO$_3$/LaMnO$_3$
%heterostructure at different magnetic fields perpendicular to c
%axis.  }
%
% \label{fig3}
%\end{figure}
%

\ \\

{\bf II. Description of the polarization catastrophe and how we can
control the superconducting properties.}

The polar discontinuity at the interface leads to the divergence of
the electrostatic potential. This leads to the appearance of the
electric field outside of the ferroelectric film and substantial
increase in total energy. In order to minimize the total energy, it is
necessary to screen this electric field. As a consequence, both the
lattice system and the energy spectrum of the current carriers are
restructured [5], and the current carrier density in a narrow
interface area is increased[5, 6]. This occurs in a self-consistent
manner, therefore, the renormalization  the energy spectrum and
increasing of carrier concentration in the interface leads to the
formation of a narrow metallic region in LCO near the interface. In
order to estimate the carrier concentration we assume that the
abrupt change of the polarization at the interface leads to the
formation of the surface screening charge with the quasi 2D density
$\sigma_S=P_n $ where $P_n$ is the component of polarization
perpendicular to the interface. If we assume that the value of
polarization at the surface is the same as in the bulk of
ferroelectric ($P_n\approx 30\  \mu $K/cm$^2 $) we obtain $\sigma_S$
= 1.875\ 10$^{14}$ cm$^{-2}$. The superconductivity in
La$_{2-x}$Sr$_{x}$CuO$_4$ is observed at $x$ = 0.05-0.26 with the
carrier concentration of 0.5 10$^{14}$ cm$^{-2}$ per CuO plane for $x$
= 0.15. Therefore, we can expect that the superconducting state will
be localized in the narrow region of LCO of the order of 1-5 nm. But
we have to underline that in real system the thickness of the
superconducting layer will be strongly dependent on selfconsistent
renormalization of all parameters of LCO near the interface.

 Note that superconducting state (SCS) on the interface depends on the direction of
polarization. When the polarization is orthogonal to the interface
SCS will be observed. On the other hand, when the polarization is
parallel to the interface the superconductivity will not be
observed. Moreover, depending on the direction of polarization (up
or down) it is possible to observe electron and hole doped
superconductivity. This is very important because it allows to
investigate electron and hole doped superconductivity under
identical conditions. Therefore, by applying the external electric
field and switching the direction of polarization from perpendicular
to parallel direction to the interface it is possible to turn the 
superconductivity at the interface on and off. On the other hand, changing
the direction of polarization by 180 degrees, when it is
perpendicular to the interface, it is possible to change the type of
carriers in superconducting state.

The indirect evidence that, at the interface of dielectric and
ferroelectric the highly conducting state is created due to
discontinuity of polarization was obtained on the
Ba$_{0.8}$Sr$_{0.2}$TiO$_3$/LaMnO$_3$ (BSTO/LMO) heterostructure
[7]. The first important observation is that highly conducting state
is observed if the polarization of the ferroelectric is
perpendicular to the interface. The effect is absent if the
polarization is parallel to the interface [7]. Second important
observation is that highly conducting state was turned off by the
applying the inhomogeneous electric field. After that the
conductivity demonstrated ordinary semiconducting behaviour. After
application of the homogeneous electric field perpendicular to the
interface the highly conducting state were recovered. Our attempts
to obtain similar results on BSTO/LCO heterostructure did not lead us
to the expected result. The main reason is that the application of
the strong inhomogeneous electric field leads to the destruction of
the sample because of electrostriction. The switching of the
superconducting state by the applying the electric field is the goal
of our future project.  But we believe that the results obtained for
BSTO/LCO heterostructure are related with the polarization
catastrophe at the interface, because both heterostructures were
obtained by using the same technique, using the same equipment and
under the same conditions.

\begin{figure}[htb]
%\centering
% \vspace{-1.0cm}
 \includegraphics[angle=0,width=1.0\linewidth]{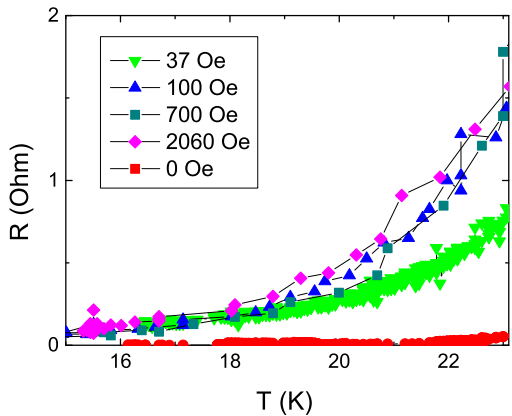}

 \vspace{-0.0cm}
% closed and the open symbols refer to FeSe$_{0.94}$ and
\caption{ The temperature dependence of
the resistance of Ba$_{0.8}$Sr$_{0.2}$TiO$_3$/La$_2$CuO$_4$
heterostructure at different magnetic fields in low temperatures
range.}
 \label{fig:Phase-diagramm}
\end{figure}

\begin{figure}[htb]
%\centering
 \vspace{-0.0cm}
\includegraphics[angle=0,width=1.0\linewidth]{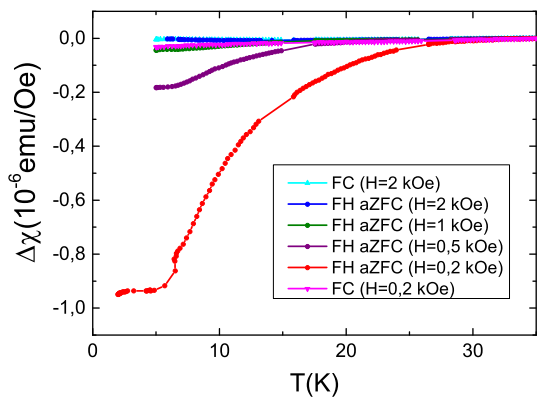}
 \vspace{-0.0cm}
\caption{ The temperature dependence of
the diamagnetic susceptibility  of
Ba$_{0.8}$Sr$_{0.2}$TiO$_3$/La$_2$CuO$_4$ heterostructure in low
temperatures range, measured in the magnetic fields 200 Oe, 500 Oe and 2000 Oe. }
 \label{fig4a}
\end{figure}
\ \\
\begin{figure}[htb]
%\centering
 \vspace{-0.0cm}
\includegraphics[angle=0,width=1.0\linewidth]{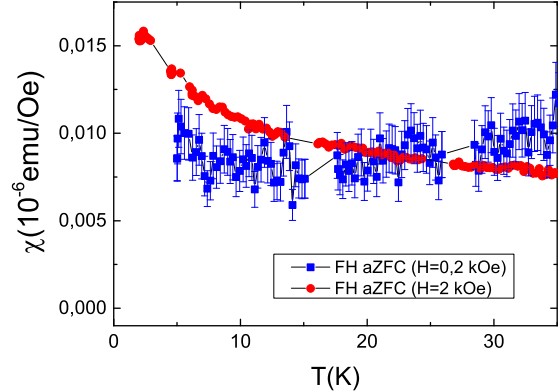}
 \vspace{-0.0cm}
\caption{ The temperature dependence of the magnetic susceptibility  of
La$_2$CuO$_4$ single crystal heated in the deposition chamber, at the same oxygen 
partial pressure and temperature,  as used during the heterostructure fabrication in low
temperatures range, measured in the magnetic fields 200 Oe, and 2000 Oe. The error bars for the field 2000 Oe are less than the size of the markers.}
 \label{fig4b}
\end{figure}
\ \\

{\bf III.Measurements of resistance and superconducting state at the interface}.
\ \\

Resistance measurement on the interface of the BSTO/LCO
heteroctructure was performed by four contacts method as shown on
Fig. 1a of the article. The electrodes were applied on the LCO
surface at the boundary with film. The electrodes were in contact
with the interface. The distance between potential electrodes was
different in the different experiments. The current flows by
different routes at different temperatures and depends on relation
between a substrate conductance and an interface conductance. At
high temperature the main current flows through substrate. Below 50
K the main current flows on interface area. The measurement of
resistance at magnetic field was performed by two different ways. In
one case the heterostructure has been cooled without magnetic field
applied and then the resistance was measured in magnetic field at
certain temperature changing the magnetic field step by step from 0
to 4000 Oe. In the other case the measurement of resistance was
carried out during cooling at the different magnetic fields (Fig. 3
of the article).

\ \\

{\bf IV. Results of the resistivity and susceptibility measurements 
in different fields. Description of the flux flow resistance}.

\ \\

The main result of the effect of external magnetic field on
superconducting state is presented in Fig.3 of the main text. These
results show that vortices start to penetrate to superconductor at
very low field. The penetration of vortices leads to nonzero resistance
due to flux flow resistance: $R_{ff}=R_0 H/H_{c2}$ ($R_0$ is the
normal state resistance). Therefore, if we take $R_0$ = 200 Ohm,
$H$=0,02-0,04 T and $H_{c2}$ = 29-81 T,we estimate $R_{ff}$=
0,05-0,26 Ohm. In Fig. 3 we present the temperature dependence of
the resistance of the BSTO/LCO heterostructure in different magnetic
fields. It is clearly seen from this figure that the effect is
stable at different temperatures. In order to demonstrate the
superconducting state in the heterostructure we measured the
zero field cooled and field cooled magnetic susceptibility in 
different magnetic fields (Fig.4). Observation of the diamagnetic
susceptibility (Fig.4) confirms the existence of superconducting
phase at the interface. A very rough estimate of the superconducting
layer thickness from diamagnetic susceptibility measurements
provides value of 4-100 nm.

In order to rule out possible diffusion of oxygen to the LCO single crystal, 
we repeated the experiments on a La$_2$CuO$_4$ single crystal (from 
the same batch) heated in the deposition chamber, at the same oxygen 
partial pressure and temperature,as used during the heterostructure fabrication.
The results of the susceptibility measurements in different magnetic fields 
are presented in Fig.5. The results suggest that the LCO single crystal does 
not have any diamagnetic response at low temperatures and therefore does not 
have any measurable superconducting phase. As it was mentioned above 
superconducting state in this samples was not detected by resistivity 
measurements as well.

\begin{figure}[htb]
%\centering
 \vspace{-0.0cm}
\includegraphics[angle=0,width=1.0\linewidth]{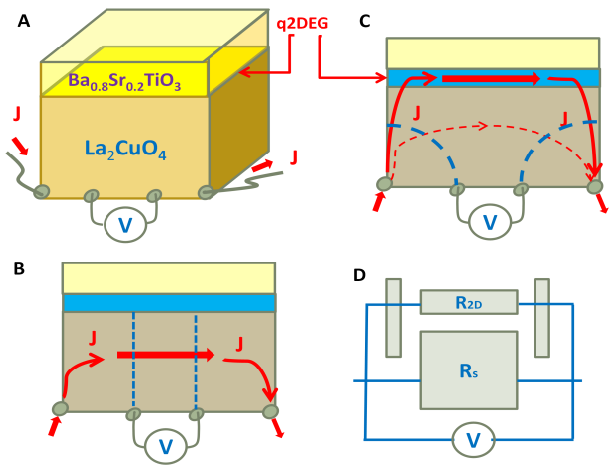}
 \vspace{-0.0cm}
\caption{ Schematic illustration of the
resistance measurement from the back side of
Ba$_{0.8}$Sr$_{0.2}$TiO$_3$/La$_2$CuO$_4$ heterostructure: contacts
layout (A), the schematic current distribution is shown by lines
with arrows for T$>$50 K (B) and T$<$50 K (C) the lines of potential
are shown by the dash lines, and scheme of the effective resistances
between potential electrodes (R$_{2D}$ is resistance of interface,
R$_S$ is resistance of substrate, the additional resistances of
substrate in the vicinity of interface are shown by vertical
parallelepipeds) (D). Below 50 K the main current flows in the
interface area (C). }
 \label{fig4}
\end{figure}

\ \\

{\bf V. Measurements of resistance from the substrate side. }

\ \\

Resistance measurement from the back side of heteroctructure was
performed as shown on Fig. 6a. The electrodes were deposited on the
LCO surface opposite to the surface with film. And electrodes were
not in contact with the interface. We believe that the current line
distributions are strongly different at different temperatures (Fig.
6b and 6c) and depend on relation between a substrate conductance
and an interface conductance. At high temperature the main current
flows through substrate. Below 50 K the main current flows on
interface area. And when the resistance is measured from the side of
substrate the temperature dependence of the heterostructure is the
same as that for the interface (Fig. 4 of the article). But
superconductivity is not observed directly because the surface of
substrate is not superconducting and the resistance is observed from
superconducting state (for T$<$30 K) and from part of substrate
(Fig. 6d). The result of the resistance measurement is presented on
Fig. 4 of the article. The conclusion is that there is no
superconductivity of the substrate surface. It means that the
interstitial oxygen does not penetrate in the LCO surface in the
interface area during the sputtering of the film.

%\end{document}
%\begin{thebibliography}{99}
%

[1] V.M. Mukhortov, Y.I. Golovko, G.N. Tolmachev and A.N. Klevtzov, Ferroelectrics \textbf{247}, 75 (2000).

[2] V. M. Mukhortov, G. N. Tolmachev, Yu. I. Golovko, A. I.
Mashchenko, Technical Physics, \textbf{43}, 1097 (1998).

[3] V. M. Mukhortov, Yu. I. Golovko, G. N. Tolmachev, A. I.
Mashchenko, Technical Physics, \textbf{44}, 1477 (1999).

[4]  A. S. Anokhin, A. G. Razumnaya, Y. I. Yuzyuk, Y. I. Golovko,
and V. M. Mukhortov, Physics of the Solid State \textbf{58}, 2027
(2016).

%[5] A. Gozar \textit{et. al.}, Nature (London) \textbf{455}, 782
%(2008).
%
%[6] A. Gozar and I. Bozovic, Physica C \textbf{521-522}, 38 (2016).

[5] V. V. Kabanov, I. I. Piyanzina, D. A. Tayurskii, and R. F.
Mamin, Phys. Rev. B \textbf{98}, 094522 (2018).

[6] X. Liu, E. Y. Tsymbal, K. M. Rabe, Phys. Rev. B \textbf{97},
094107 (2018).

[7] D. P. Pavlov {\it et. al.}, JETP Lett. \textbf{106}, 460 (2017).

%\end{thebibliography}

\end{document}